\documentclass[aps,twocolumn,superscriptaddress,showpacs]{revtex4-1}

\usepackage{graphicx}% Include figure files
\usepackage{dcolumn}% Align table columns on decimal point
\usepackage{bm}% bold math
\usepackage{amsmath}
\usepackage{color}
\usepackage{ulem}
\usepackage{threeparttable}
\usepackage{enumerate}
\usepackage{paralist}

\begin{document}

\title{Possible quantum spin liquid state of CeTa$_7$O$_{19}$}

\author{N. Li}
\thanks{These authors contributed equally to this work.}
\affiliation{Anhui Provincial Key Laboratory of Magnetic Functional Materials and Devices, Institutes of Physical Science and Information Technology, Anhui University, Hefei, Anhui 230601, People's Republic of China}

\author{A. Rutherford}
\thanks{These authors contributed equally to this work.}
\affiliation{Department of Physics and Astronomy, University of Tennessee, Knoxville, Tennessee 37996, USA}

\author{Y. Y. Wang}
\affiliation{Anhui Provincial Key Laboratory of Magnetic Functional Materials and Devices, Institutes of Physical Science and Information Technology, Anhui University, Hefei, Anhui 230601, People's Republic of China}

\author{H. Liang}
\affiliation{Anhui Provincial Key Laboratory of Magnetic Functional Materials and Devices, Institutes of Physical Science and Information Technology, Anhui University, Hefei, Anhui 230601, People's Republic of China}

\author{Y. Zhou}
\affiliation{Anhui Provincial Key Laboratory of Magnetic Functional Materials and Devices, Institutes of Physical Science and Information Technology, Anhui University, Hefei, Anhui 230601, People's Republic of China}

\author{Y. Sun}
\affiliation{Anhui Provincial Key Laboratory of Magnetic Functional Materials and Devices, Institutes of Physical Science and Information Technology, Anhui University, Hefei, Anhui 230601, People's Republic of China}

\author{D. D. Wu}
\affiliation{Anhui Provincial Key Laboratory of Magnetic Functional Materials and Devices, Institutes of Physical Science and Information Technology, Anhui University, Hefei, Anhui 230601, People's Republic of China}

\author{P. F. Chen}
\affiliation{Anhui Provincial Key Laboratory of Magnetic Functional Materials and Devices, Institutes of Physical Science and Information Technology, Anhui University, Hefei, Anhui 230601, People's Republic of China}

\author{Q. J. Li}
\affiliation{School of Physics and Optoelectronics, Anhui University, Hefei, Anhui 230061, People's Republic of China}

\author{H. Wang}
\affiliation{Department of Chemistry, Michigan State University, East Lansing, Michigan 48824, USA}

\author{W. Xie}
\affiliation{Department of Chemistry, Michigan State University, East Lansing, Michigan 48824, USA}

\author{E. S. Choi}
\affiliation{National High Magnetic Field Laboratory, Florida State University, Tallahassee, FL 32310, USA}

\author{S. Z. Zhang}
\affiliation{Los Alamos National Laboratory, Los Alamos, New Mexico 87545, USA}

\author{M. Lee}
\affiliation{Los Alamos National Laboratory, Los Alamos, New Mexico 87545, USA}

\author{H. D. Zhou}
\email{hzhou10@utk.edu}
\affiliation{Department of Physics and Astronomy, University of Tennessee, Knoxville, Tennessee 37996, USA}

\author{X. F. Sun}
\email{xfsun@ahu.edu.cn}
\affiliation{Anhui Provincial Key Laboratory of Magnetic Functional Materials and Devices, Institutes of Physical Science and Information Technology, Anhui University, Hefei, Anhui 230601, People's Republic of China}

\date{\today}

\begin{abstract}

CeTa$_7$O$_{19}$ is a recently found two-dimensional triangular lattice antiferromagnet without showing magnetic order. We grew high-quality CeTa$_7$O$_{19}$ single crystals and studied the low-temperature magnetic susceptibility, specific heat and thermal conductivity. The dc magnetic susceptibility and magnetization reveal its nature of effective spin-1/2, easy axis anisotropy, and antiferromagnetic spin coupling. The ultralow-temperature ac susceptibility and specific heat data indicate the absence of any phase transition down to 20 mK. The ultralow-temperature thermal conductivity ($\kappa$) at zero magnetic field exhibits a non-zero residual term $\kappa_0/T =$ 0.0056 W/K$^2$m. Although the magnetic field dependence of $\kappa$ is rather weak, the 14 T thermal conductivity shows an essential zero residual term. All these results point to a possible ground state of quantum spin liquid.

\end{abstract}

%\pacs{66.70.-f, 75.47.-m, 75.50.-y}
%66.70.-f Nonelectronic thermal conduction and heat-pulse propagation in solids
%75.47.-m Magnetotransport phenomena; materials for magnetotransport
%75.50.-y Studies of specific magnetic materials

\maketitle

\section{INTRODUCTION}

Two-dimensional (2D) triangular lattice antiferromagnet (TAF) with spin-1/2 is the simplest geometrically frustrated system with strong quantum spin fluctuations, which can exhibit exotic quantum magnetisms including the celebrated quantum spin liquid (QSL) \cite{Nature464, RevModPhys89, RepPeogPhys80, Science367}. Actually, Anderson first proposed the resonating valence bond (RVB) model to describe this novel quantum disorder state based on Heisenberg TAFs \cite{MaterResBull8}. In real materials, the organic TAFs $\kappa$-(BEDTTTF)$_2$Cu$_2$(CN)$_3$ and EtMe$_3$Sb[Pd(dmit)$_2$]$_2$ with $S =$ 1/2 are the early identified QSL candidates \cite{Science372, Science328}. For inorganic materials, some QSL candidates have also been found in TAFs, including YbMgGaO$_4$, $ARCh_2$ ($A =$ alkali or monovalent ions, $R =$ rare earth ions, $Ch =$ chalcogenides), Na$_2$BaCo(PO$_4$)$_2$, Pr$M$Al$_{11}$O$_{19}$ ($M =$ Zn, Mg), Ba$_6$$R_2$Ti$_4$O$_{17}$ ($R =$ rare earth ions), YbZn$_2$GaO$_5$, NaRuO$_2$, and TbInO$_3$ etc. \cite{SciRep5, CPL35, PNAS116, JAlloysCompd, JMaterChem, arXiv08937, arXiv20040, NatPhys19, NatPhys15} However, the inevitable disorder effects such as vacancies, impurities, lattice distortion, and stacking faults in real materials induce serious difficulties in accurately characterizing the real ground sate in theory and experiments. For example, there is a site mixing between nonmagnetic Mg$^{2+}$ and Ga$^{3+}$ ions in YbMgGaO$_4$, causing the controversy on the ground state and the low-energy magnetic excitations \cite{Nature540, PhysRevLett117, PhysRevLett119}.

NdTa$_7$O$_{19}$ was recently proposed as a rare QSL candidate with Ising anisotropy \cite{NatMater416}, based on the magnetization, electron spin resonance and muon spin relaxation results. This material has a hexagonal crystal structure with the space group $P\overline{6}c2$. In this compound, the Nd$^{3+}$ ions with effective 1/2-spin form a 2D triangular lattice without structural disorder. Later, Wang {\it et al.} synthesized various $R$Ta$_7$O$_{19}$ ($R =$ Pr, Sm, Eu, Gd, Dy, Ho) which all do not order at low temperatures down to 2 K \cite{Wang}. Very recently, another two sister materials CeTa$_7$O$_{19}$ and YbTa$_7$O$_{19}$ were also synthesized and found to have possible QSL ground state \cite{2411.18045}. CeTa$_7$O$_{19}$ is also a Ising-like TAF with very weak exchange interaction ($\sim$ 0.22 K), while YbTa$_7$O$_{19}$ has an in-plane magnetic anisotropy. It is notable that all these previous studies demonstrated that $R$Ta$_7$O$_{19}$ have perfect triangular spin lattice and would be a good platform to probe the possible QSL in TAFs. Till now, the single crystal samples are not available for most members of this family, which prevents more characterizations on the nature of ground states.

Three experimental hallmarks have been widely accepted as evidence for the QSL state and the associated spinon excitations, including a broad continuous magnetic intensity in the inelastic neutron scattering spectrum \cite{Nature540, NP13-117}, a large magnetic specific heat with power-law temperature dependence \cite{NP4-459, PRL95-036403}, and a non-zero residual thermal conductivity $\kappa_0/T$ at zero-Kelvin limit \cite{Science328, NC4949, NC4216, PhysRevsearch013099}. In this work, we grew high-quality CeTa$_7$O$_{19}$ single crystal and studied its magnetic ground state by using the ultralow-temperature thermal conductivity, as well as the magnetic susceptibility and specific heat measurements. Besides the absence of magnetic order transition indicated by the ultralow-temperature ac susceptibility and specific heat, a non-zero $\kappa_0/T$ was observed. These results suggest a possible QSL ground state of CeTa$_7$O$_{19}$.

\section{EXPERIMENTS}

CeTa$_7$O$_{19}$ single crystals were grown by using the flux method. The mixture of CeO$_2$, Ta$_2$O$_5$, and V$_2$O$_5$ with molar ratio 0.25:0.1:0.65 was loaded in a Pt crucible with cover. The mixture was annealed at 1200 $^{\circ}$C for 40 hours and then cooled down to 900 $^{\circ}$C with 0.5 $^{\circ}$C/hour rate, and thereafter was cooled to room temperature by air quench. The flux was dissolved in hot HCl acid and crystals with plate shape were recovered. The Laue back diffraction measurement was used to determine the crystalline orientation. The single crystal X-ray diffraction (SCXRD) measurement was performed using a XtalLAB Synergy, Dualflex, Hypix single crystal X-ray diffractometer with Mo K$_{\alpha}$ radiation ($\lambda$ = 0.71073 \AA). The structure was solved and refined using the Bruker SHELXTL Software Package.

The dc magnetic susceptibility between 2 and 300 K were measured using the SQUID-VSM (Quantum Design Magnetic Property Measurement System, MPMS) in magnetic field up to 7 T. The ac susceptibility was measured using the conventional mutual inductance technique (with a combination of ac current source and a lockin amplifier) at SCM1 dilution fridge magnet of the National High Magnetic Field Laboratory, Tallahassee. The high field magnetization was measured in the pulsed field facility at Los Alamos National Laboratory by detecting the field-induced voltage across a pick-up coil made of high purity Cu. The specific heat was measured by relaxation technique using the Physical Property Measurement System (PPMS) (DynaCool, Quantum Design) with a dilution insert. The thermal conductivity was measured by using ``one heater, two thermometers" technique in a dilution refrigerator equipped with a 14 T magnet \cite{PhysRevB184423, NC4949, NC4216}. The heat current was along the longest direction ($a$ axis) and the external magnetic fields were applied along the $a$ axis or the $c$ axis.

\section{RESULTS AND DISCUSSIONS}

\begin{figure}
\includegraphics[clip,width=8.5cm]{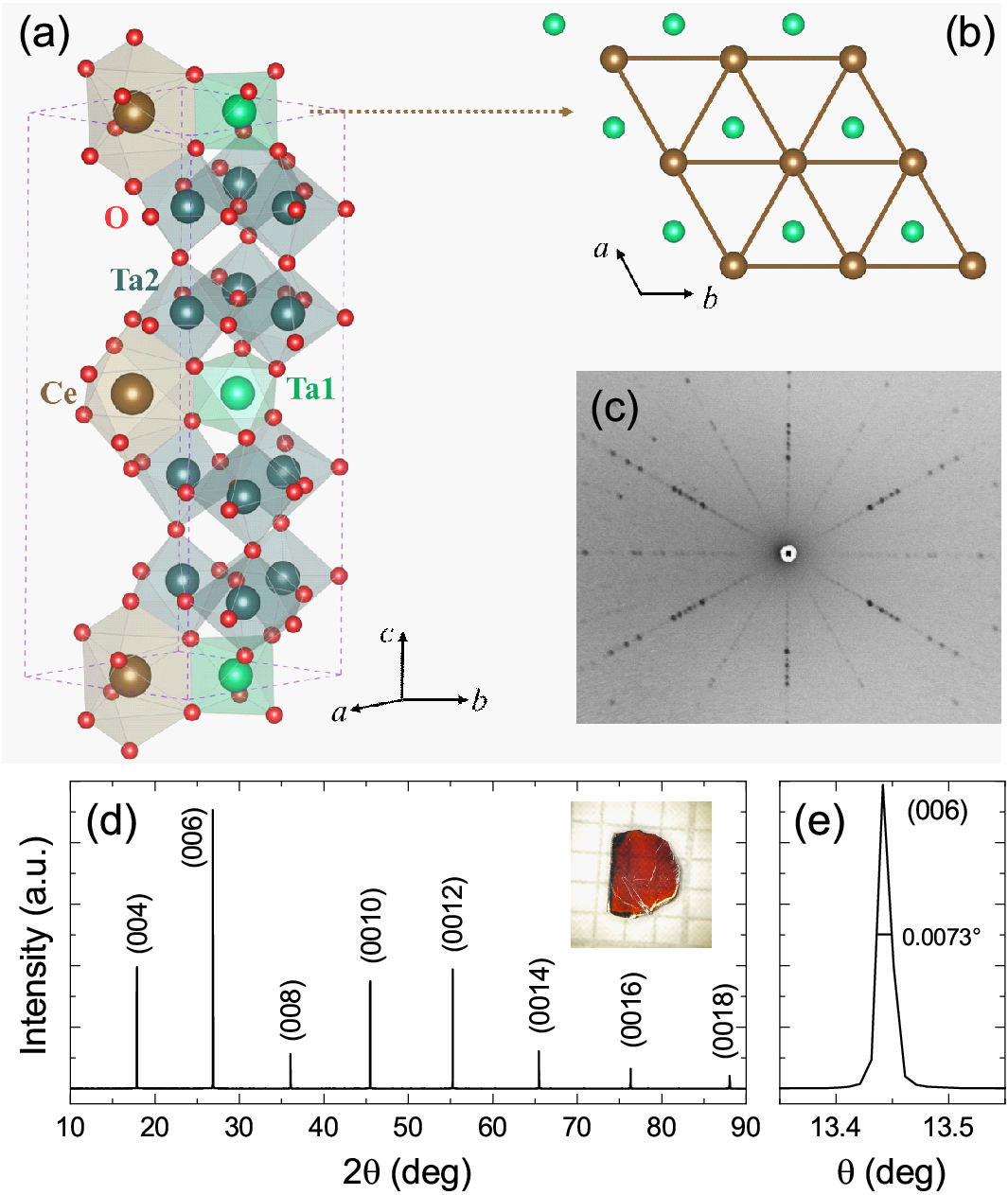}
\caption{(a,b) The schematic crystal structure of CeTa$_7$O$_{19}$, where Ce$^{3+}$ ions form into an ideal triangular layer in the $ab$ plane and these triangular layers are stacked along the $c$ axis with the nearest-neighbor interlayer distance of 9.9915 \AA. (c) The X-ray Laue back diffraction photo for the $ab$ plane. (d,e) X-ray diffraction pattern of (00$l$) plane and the rocking curve of the (006) peak. The full width at the half maximum (FWHM) of the rocking curve is shown in panel (e). The inset to panel (d) shows the photograph of CeTa$_7$O$_{19}$ single crystal.}
\end{figure}

The refinement of the SCXRD data confirms that CeTa$_7$O$_{19}$ is isostructural to NdTa$_7$O$_{19}$ with Ce$^{3+}$ ions forming the 2D triangular lattice, as shown in Figs. 1(a) and 1(b). The refinement results are listed in Table I and II. Our CeTa$_7$O$_{19}$ single crystals are brown colored with typical size of 2--3 mm, as shown in the inset to Fig. 1(d). The good crystallinity was confirmed by the X-ray back reflection Laue photograph, as shown in Fig. 1(c). Furthermore, Fig. 1(d) shows the X-ray diffraction pattern of single crystals, which was found to be the (00$l$) diffraction. No impurity phase was detected. The X-ray rocking curve of the (006) peak is very narrow (the full width of the half maximum is only 0.0073$^\circ$), which also indicates the good crystallinity.

\begin{table*}[htbp]
	\caption{The crystal structure and refinement for CeTa$_7$O$_{19}$ at room temperature (Mo K$\alpha$ radiation).Values in parentheses are estimated standard deviation from refinement.} % title of Table
	\centering % used for centering table
	\begin{tabular}{c c} % centered columns (5 columns)
		\hline\hline %inserts double horizontal lines
	    $\textbf{Chemical}$ $\mathbf{Formula}$ & $\textbf{CeTa$_7$O$_{19}$}$ \\ % inserts table
		%heading
		\hline % inserts single horizontal line
        Formula weight & 1710.77 g/mol \\ % inserting body of the table
        Space Group & $P-6c2$ \\
		Unit cell dimensions & $a =$ 6.2371(2) \AA \\
        ~\ & $c =$ 19.9830(7) \AA \\
	    Volume & 673.23(4) \AA$^3$ \\
        $Z$ & 2 \\
		Density(calculated) & 8.439 g/cm$^3$ \\
		Absorption coefficient & 60.043 mm$^{-1}$ \\
        F(000) & 1442.0 \\
        2$\theta$ range & 7.544 to 82.132$^{\circ}$ \\
        Reflections collected & 8108 \\
        Independent reflections & 1489 [$R\rm_{int} =$ 0.0888] \\
        Refinement method & Full-matrix least-squares on F$^2$ \\
        Data/restraints/parameters & 1489 / 0 / 45 \\
        Final R indices & $R_1$ ($I >$ 2$\sigma$($I$)) = 0.0341; $wR_2$ ($I >$ 2$\sigma$($I$)) = 0.0763 \\
        ~\ & $R_1$(all) = 0.0385; $wR_2$ (all) = 0.0789 \\
        Largest diff. peak and hole & +4.01 e/\AA$^3$ and -3.08 e/\AA$^3$ \\
        R.M.S. deviation from mean & 0.955 e/\AA$^3$ \\
        Goodness-of-fit on F$^2$ & 1.046 \\

		\hline %inserts single line
	\end{tabular}
	\label{table:nuclear} % is used to refer this table in the text
\end{table*}

\begin{table*}[htbp]
	\caption{Atomic coordinates and equivalent isotropic atomic displacement parameters (\AA$^2$) of CeTa$_7$O$_{19}$. (U$_{eq}$ is defined as one third of the trace of the orthogonalized U$_{ij}$ tensor.)} % title of Table
	\centering % used for centering table
	\begin{tabular}{c c c c c c c} % centered columns (5 columns)
		\hline\hline %inserts double horizontal lines
		$\textbf{CeTa$_7$O$_{19}$}$ & $\textbf{Wyck.}$ & $\textbf{x}$ & $\textbf{y}$ & $\textbf{z}$ & $\textbf{Occ.}$ & $\textbf{U$_\text{eq}$}$ \\ [0.5ex] % inserts table
		%heading
		\hline % inserts single horizontal line
		Ta$_1$ & 2$a$ & 0 & 0 & 0 & 1 & 0.006(2) \\ % inserting body of the table
        Ta$_2$ & 12$l$ & 0.02593(2) & 0.33193(12) & 0.15613(2) & 1 & 0.007(1) \\
        Ce$_1$ & 2$c$ & 1/3 & 2/3 & 0 & 1 & 0.008(1) \\
        O$_1$ & 12$l$ & 0.42017(19) & 0.33583(20) & 0.15360(3) & 1 & 0.013(1) \\
        O$_2$ & 6$k$ & 0.08257(18) & 0.37653(20) & 1/4 & 1 & 0.010(2) \\
        O$_3$ & 12$l$ & 0.04227(12) & 0.27983(12) & 0.05690(3) & 1 & 0.009(1) \\
        O$_4$ & 4$g$ & 0 & 0 & 0.16830(6) & 1 & 0.007(2) \\
        O$_5$ & 4$h$ & 1/3 & 2/3 & 0.13250(7) & 1 &  0.009(2) \\
		\hline %inserts single line
	\end{tabular}
	\label{table:nuclear2} % is used to refer this table in the text
\end{table*}

\begin{figure}
\includegraphics[clip,width=8.5cm]{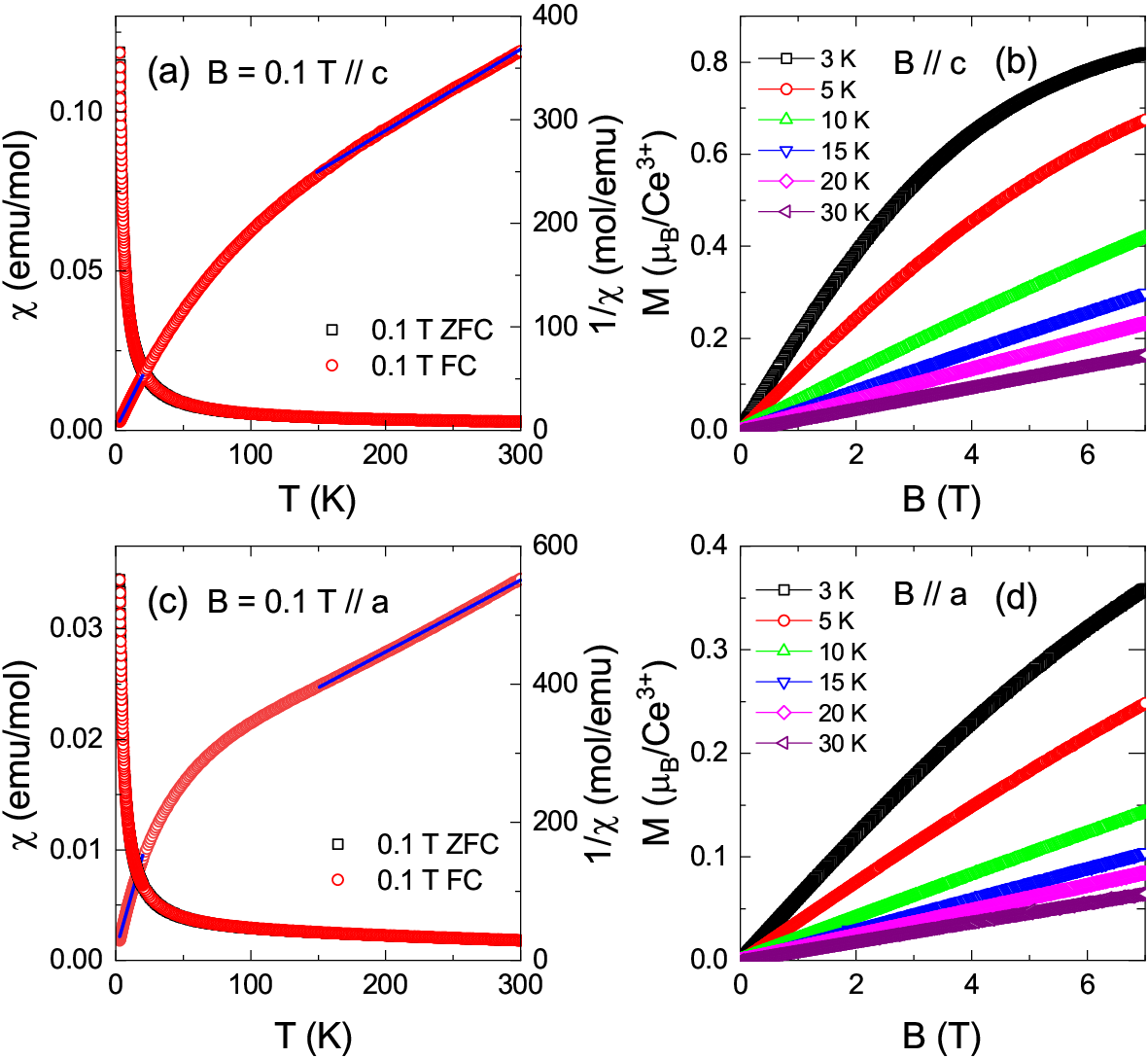}
\caption{Temperature dependence of magnetic susceptibility $\chi$ and inverse magnetic susceptibility 1/$\chi$ of CeTa$_7$O$_{19}$ single crystals with $B \parallel c$ (a) or $B \parallel a$ (c). The solid blue lines indicate the linear fit of the inverse magnetic susceptibility 1/$\chi(T)$ curves. Isothermal magnetization of CeTa$_7$O$_{19}$ single crystals at selected temperatures with $B \parallel c$ (b) or $B \parallel a$ (d).}
\label{Res}
\end{figure}

Figures 2(a) and 2(c) present the temperature dependence of magnetic susceptibility $\chi$ measured in $B =$ 0.1 T along the $c$ and $a$ axis, respectively. The $\chi(T)$ increases upon lowering temperature and no visible anomaly can be observed down to 2 K, indicating the absence of long-range magnetic order. In addition, the zero-field-cooling (ZFC) and field-cooling (FC) data are identical, which indicates there is no spin freezing. The temperature dependence of $\chi$ does not follow Curie-Weiss law well in the whole temperature range, although the high-temperature and low-temperature data can be separately fitted. The Curie-Weiss fit to the low-temperature data gives $\mu_{eff} =$ 1.755 $\mu_B$, $\theta_{CW} = -$0.469 K for $B \parallel c$ and $\mu_{eff} =$ 1.169 $\mu_B$, $\theta_{CW} = -$1.383 K for $B \parallel a$, respectively. These results indicate the dominant antiferromagnetic interactions. Figures 2(b) and 2(d) show the isothermal magnetization curves at selected temperatures with the external magnetic fields along the $c$ and $a$ axis, respectively. The data clearly suggest easy axis anisotropy since the magnetization reaches around 0.8 $\mu_B$ at $B =$ 7 T for $B \parallel c$, but only 0.35 $\mu_B$ at $B =$ 7 T for $B \parallel a$.

The recent crystal electric field (CEF) studies \cite{2411.18045} on polycrystalline CeTa$_7$O$_{19}$ samples reveal that (i) the first excited CEF doublet lies at a very large energy gap 42.9 meV above the ground state and therefore, at low temperature, the ground state of Ce$^{3+}$ ions could be treated as effective spin-1/2; (ii) the powder averaged saturation is around 0.82 $\mu_B$ and $\mu_{eff} =$ 1.42 $\mu_B$; (iii) the system is Ising like. All these observations are consistent with our magnetic properties listed above for crystalline samples.

\begin{figure}
\includegraphics[clip,width=8.5cm]{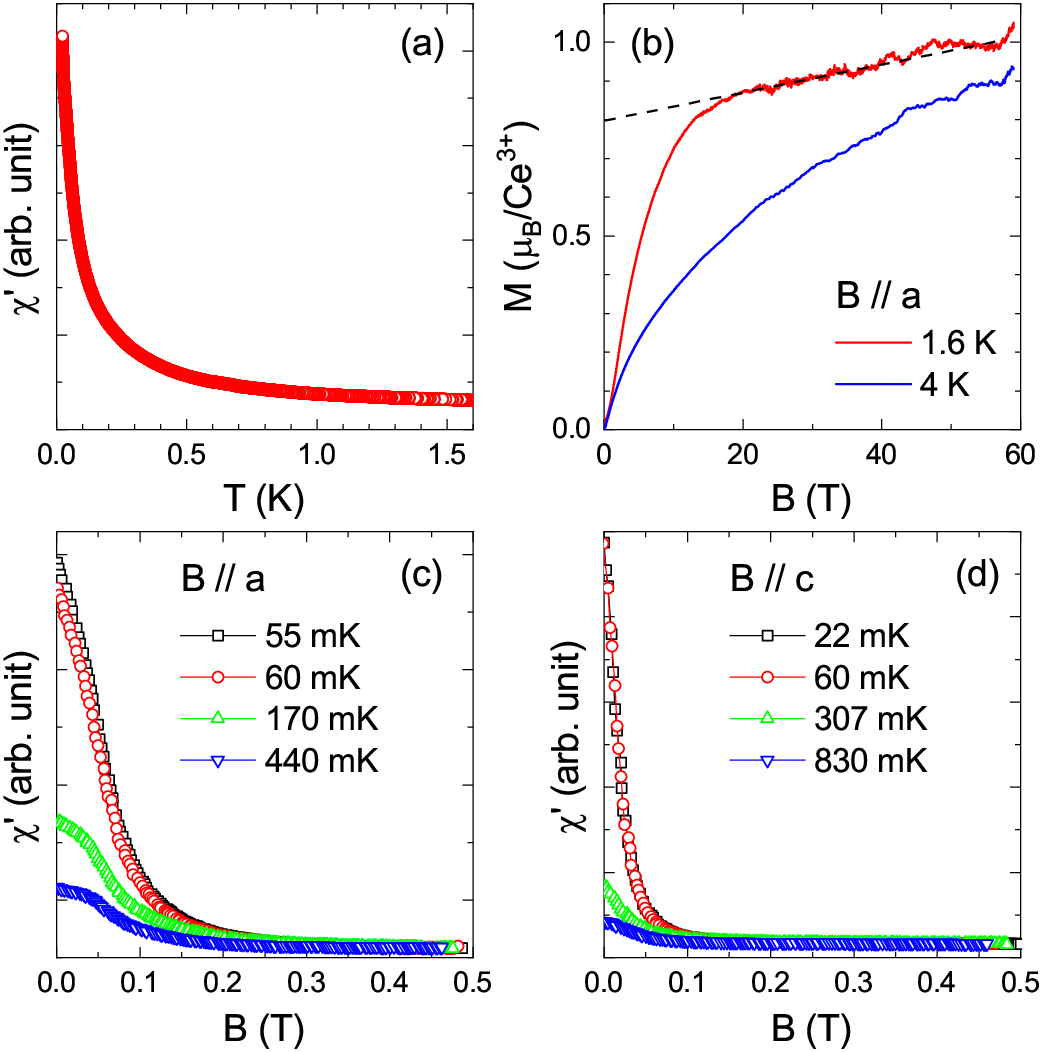}
\caption{(a) Ultralow-temperature ac magnetic susceptibility of CeTa$_7$O$_{19}$ single crystal with frequency 220 Hz and 0.5 Oe ac field. (b) High-field magnetization curves with the magnetic field along the $a$ axis. The dashed line shows the Van Vleck paramagnetic term. (c, d) The field dependence of ac susceptibility at different temperatures for $B \parallel a$ and $B \parallel c$.}
\end{figure}

Figure 3(a) shows the ultralow-temperature ac magnetic susceptibility of CeTa$_7$O$_{19}$ single crystal measured at zero dc field. The data show that there is no magnetic transition at low temperatures down to 20 mK. Figure 3(b) shows the high-field magnetization curves with the magnetic field along the $a$ axis. At 1.6 K, after a linear subtraction by considering the Van Vleck paramagnetic contribution, the system saturates around 15 T with saturation moment $\sim$ 0.8 $\mu_B$. This again confirms the easy axis nature of CeTa$_7$O$_{19}$ since along the $c$ axis, the system reaches 0.8 $\mu_B$ at much lower field around 7 T. Moreover, we measured the magnetic field dependence of ac susceptibility at ultralow temperatures for $B \parallel a$ and $B \parallel c$ to detect the possible field-induced transitions. As shown in Figs. 3(c) and 3(d), there is no any field-induced transition up to 0.5 T. Apparently, the magnetization behavior is quite simple in CeTa$_7$O$_{19}$, which is different from some other QSL candidates with triangular lattice \cite{NC4949, NC4216, PRB224414}.

\begin{figure}
\includegraphics[clip,width=8cm]{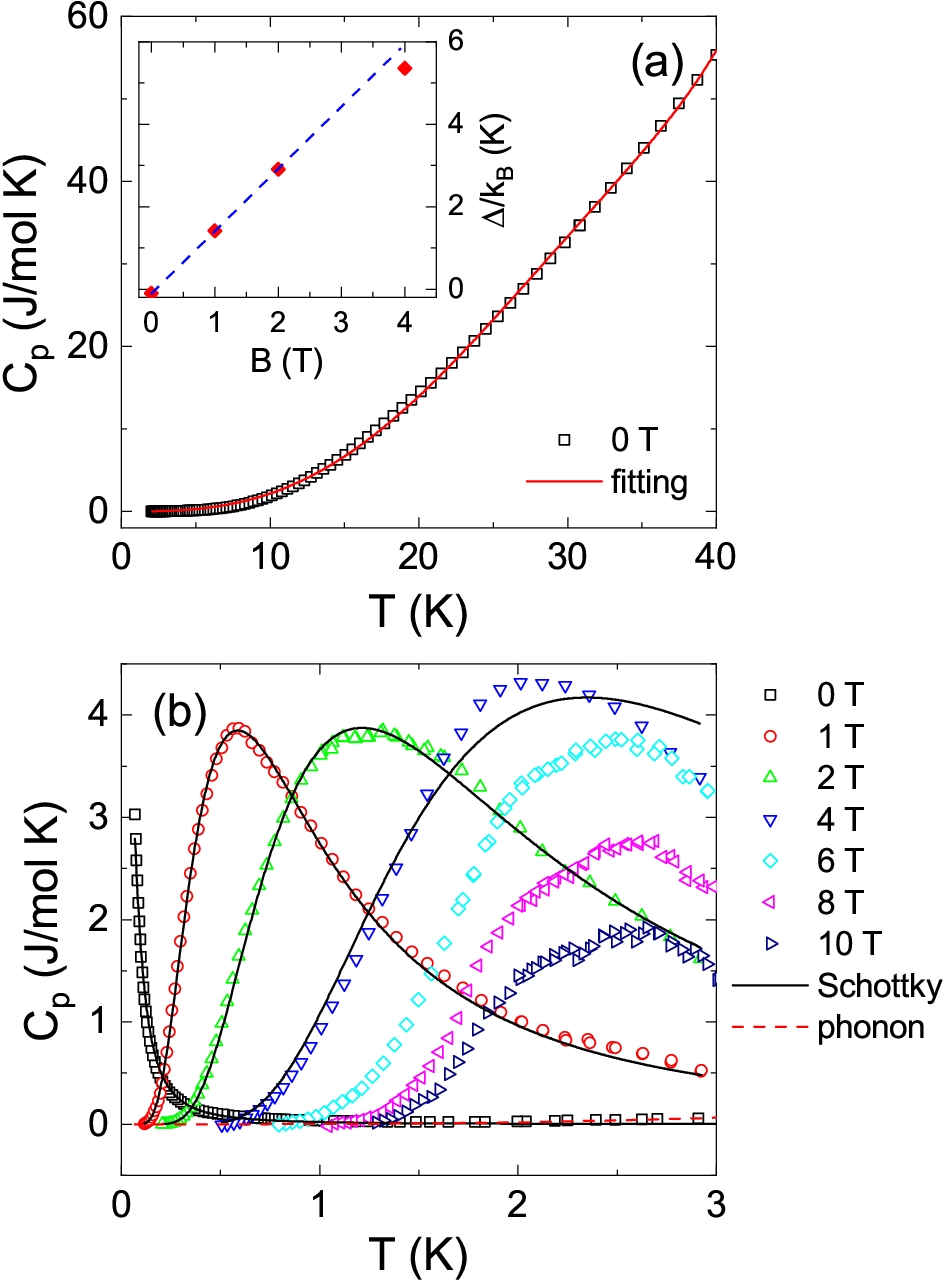}
\caption{(a) Zero-field specific heat of CeTa$_7$O$_{19}$ single crystal at 2--40 K. The solid line is the fitting result of high-temperature phononic specific heat. (b) Ultralow-temperature (down to 70 mK) specific heat under different magnetic fields with $B \parallel c$. The dashed line shows the phonon term and the solid lines show the fittings of the Schottky term. The inset shows the fitting parameter $\Delta$ in the two-level Schottky formula.}
\label{SH}
\end{figure}

Figure 4 shows the specific heat of CeTa$_7$O$_{19}$ single crystal under various magnetic fields. Clearly, there is no phase transition at low temperatures down to 70 mK, which demonstrates the disordered ground state of this material. At very low temperatures, the zero-field data show strong upturn which is likely due to the Schottky contribution of Ce$^{3+}$. With applying magnetic field along the $c$ axis, a broad peak shows up and moves to higher temperatures, which is also a typical Schottky behavior.

It can be seen that the magnetic contributions to the specific heat are significant only at very low temperatures. We can fit the phonon specific heat for the zero-field data at $T >$ 2 K. It is known that in the temperature range 0.02 $< T / \Theta_D <$ 0.1 ($\Theta_D$ is the Debye temperature), one can use the low-frequency expansion of the Debye function, $C = \beta T^3 + \beta_5 T^5+\beta_7 T^7$, where $\beta$, $\beta_5$ and $\beta_7$ are temperature-independent coefficients \cite{Tari}. As shown in Fig. 4(a), this formula gives a good fitting to the experimental data, with the fitting parameters $\beta = 2.30 \times 10^{-3}$ J/K$^4$mol, $\beta_5 = -1.56 \times 10^{-6}$ J/K$^6$mol and $\beta_7 = 4.20 \times 10^{-10}$ J/K$^8$mol. At very low temperatures, the $T^5$- and $T^7$- terms are negligible and the phonon specific heat shows a well-known $T^3$ dependence with the coefficient of $\beta$. As shown in Fig. 4(b), the phonon specific heat is much smaller than the experimental data at $T <$ 2 K. Thus, the very low temperature specific heat under various fields are dominantly the magnetic contributions. We further use the two-level Schottky formula to fit these data,
\begin{equation}
C_{sch} = nR \left(\frac{\Delta}{k_{B} T}\right)^{2} \frac{e^{\left(\Delta/k_{B}T\right)}}{\left[1+e^{\left(\Delta/k_{B}T\right)}\right]^{2}},
\end{equation}
where $n$ is the concentration of Schottky centers, $R$ is the universal gas constant, $k_B$ is the Boltzmann constant, and $\Delta$ is the gap of the Zeeman splitting of the Kramers doublet ground state of Ce$^{3+}$ ions \cite{Tari}. As shown in Fig. 4(b), the fitting are quite good for low-field data but fails at high magnetic fields. The fitting parameter $\Delta$ increases linearly with magnetic field up to 2 T, in consistent with the Zemman effect. That is, this energy gap is the splitting of the ground-state doublet of Ce$^{3+}$ crystal field levels. However, the $\Delta$ starts to deviate from linearity at 4 T. Furthermore, above 6 T, the position of peak does not change with increasing magnetic field. It is notable that the behavior of present magnetic specific heat is not common in other Ce$^{3+}$ compounds or QSL candidates. The magnetic specific heat of those materials often exhibits broad peak even in zero field and the peak gradually shifts to higher temperature with increasing magnetic field. Usually, this peak is described as the onset of coherent quantum fluctuations or spin correlations instead of the Schottky anomaly \cite{NatPhys1052, NatPhys546, PRB106-094425, arXiv15957}. In some cases, the data showed that orphan magnetic moments trapped in defects or the crystal-field excitations of rare-earth ions cause a Schottky-like anomaly and one can also use the simple two-level Schottky model to fit such kind of data \cite{PRB98-174404, PRX9-031005, PRB106-174406, arXiv07350, PRB109-224429, arXiv13070}. For a comparison, the QSL candidate Ce$_2$Zr$_2$O$_7$ has a broad peak of magnetic specific heat in zero field, which shifts to higher temperature with increasing the external magnetic field \cite{NatPhys1052}. The two-level Schottky model fitting parameter $\Delta$ also starts to deviate from linear field dependence at 4 T in Ce$_2$Zr$_2$O$_7$. Based on this analysis, the authors of this work actually ruled out the case that the existence of defects causes the broad peak of magnetic specific heat. Applying magnetic field to QSL candidates can sometimes induce phase transitions, which can be probed by the field dependence of specific heat, as seen in YbMgGaO$_4$ and NaYb$Ch_2$ \cite{NC4949, PRB224414}. However, for CeTa$_7$O$_{19}$ the values of specific heat are very small at $B >$ 6 T and $T <$ 1 K, making it difficult to detect the possible phase transitions from the specific heat data.

\begin{figure}
\includegraphics[clip,width=6.5cm]{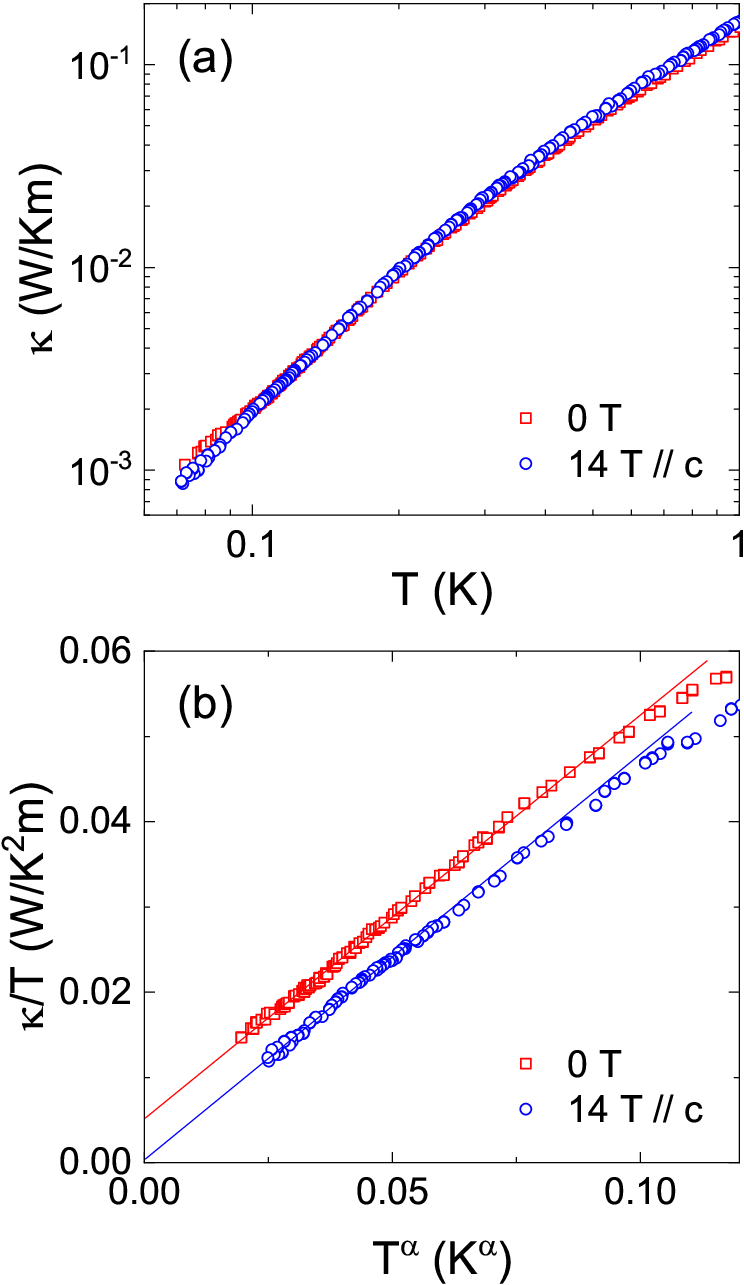}
\caption{(a) Low-temperature thermal conductivity of the CeTa$_7$O$_{19}$ single crystal in zero field and 14 T along the $c$ axis. (b) Ultralow-temperature data plotted with $\kappa/T$ vs $T^{\alpha}$ ($\alpha =$ 1.5 and 1.4 for 0 T and 14 T, respectively). The solid lines are the linear fittings to the data.}
\label{kappa}
\end{figure}

Figure 5(a) shows the low-temperature thermal conductivity of the CeTa$_7$O$_{19}$ single crystal in 0 T and 14 T with $B \parallel c$. As one can seen, the 14 T thermal conductivity is smaller than that in zero field at ultralow temperatures and is slightly larger at high temperatures. That is, the external magnetic field has rather weak effect on the thermal conductivity. We further analyzed the data at very low temperatures, as shown in Fig. 5(b). The $\kappa(T)$ curve does not show the standard $T^3$ behavior of phonon thermal conductivity at very low temperatures. This indicates that either heat carriers other than phonons or some peculiar phonon scattering effect are involved, since at subKelvin temperatures the common phonon scatterings by crystal imperfections are smeared out \cite{Oxford, PhysRev176501}. Actually, as shown in Fig. 5(b), the lowest-temperature data show a good linear behavior in the $\kappa/T$ vs $T^{1.5}$ plot. Using the formula $\kappa/T = a + bT^{1.5}$ to fit the data at $T <$ 200 mK, we get the residual term $a$ or $\kappa_0/T =$ 0.0056 W/K$^2$m for the zero-field data. Here, the two terms represent contributions from the itinerant fermionic excitations and phonons, respectively. With applying 14 T magnetic field along the $c$ axis, the data also show a linear behavior in the $\kappa/T$ vs $T^{1.4}$ plot at very low temperatures and the linear fitting gives an essential zero residual term, which is in the expectation that high magnetic field suppresses the fermionic excitations.

The non-zero $\kappa_0/T$ term at zero-Kelvin limit is often considered one of key evidences for a QSL ground state and has been extensively studied in recent years. However, the residual thermal conductivity remains a topic of debate in certain QSL candidates, such as EtMe$_3$Sb[Pd(dmit)$_2$]$_2$, YbMgGaO$_4$, and $1T$-TaS$_2$ and is mainly related to the influence of unavoidable disorders \cite{Science328, NC4949, PhysRevsearch013099, PRL247204, PRL267202, PRB081111, CPL127501}. In this work, the $\kappa_0/T$ in zero field is in the same order of magnitude with those of some quantum magnets having itinerant fermionic excitations \cite{NC4949, NC4216, PhysRevB184423}. The above thermal conductivity results indicate that in zero field there are likely itinerant spinons. Combined with the magnetic susceptibility and specific heat data, the thermal conductivity results point to a possible gapless QSL ground state of CeTa$_7$O$_{19}$.

We can estimate the mean free path ($l_s$) of the spinon excitations by using the formula \cite{Science328}
\begin{equation}
{\frac{\kappa_0}{T} = \frac{1}{3}\frac{C_s}{T} v_s l_s = \frac{\pi k_B^2}{9\hbar} \frac{l_s}{ad}}.
\end{equation}
Here, $C_s$, $v_s$ and $l_s$ are the specific heat, velocity and mean free path of spinons, respectively, and $C_s$ is assumed to have a linear temperature dependence. The $a$ ($\sim$ 6.2371 \AA) and $d$ ($\sim$ 9.9915 \AA) are nearest-neighbor intralayer and interlayer Ce-Ce distance, respectively. The spin exchange $J$ is about 0.5 K, referring to the Curie-Weiss temperature. From the observed $\kappa_0/T$ = 0.0056 W/K$^2$m, the $l_s$ is obtained as 110.8 \AA, indicating that the excitations are mobile to a distance 18 times as long as the inter-spin distance without being scattered. In addition, the spinon specific heat is about $C_s = 4.80T$ J/mol K, which is much smaller than the Schottky term and is difficult to be separated from the experimental data.

\begin{figure}
\includegraphics[clip,width=6.5cm]{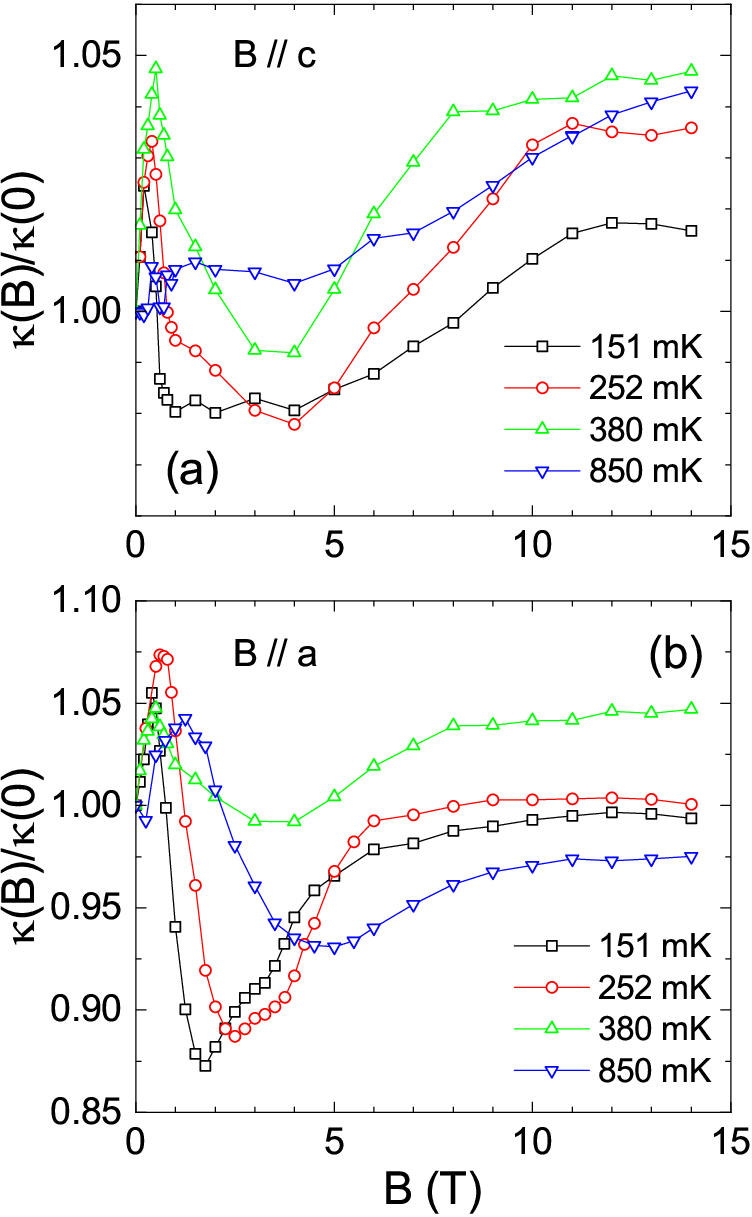}
\caption{Magnetic field dependence of thermal conductivity of the CeTa$_7$O$_{19}$ single crystal at different temperatures.}
\end{figure}

Figure 6 shows the magnetic field dependence of thermal conductivity at different temperatures. In general, the magnetic field changes the $\kappa$ rather weakly and the $\kappa(B)$ behaves similarly for different field directions. There are two features in these curves: a small peak and a broad valley at low fields and intermediate fields, respectively. It is known that the sharp peak or sharp minimum on $\kappa(B)$ curves are usually associated with the magnetic transitions, as observed in many quantum magnetic materials \cite{NC4949, NC4216, PhysRevB184423, Li2, Sun_DTN, Zhao_IPA}. However, the magnetization data does not indicate any field-induced transition in CeTa$_7$O$_{19}$. Therefore, it is more likely that this low-field small peak is due to two competitive effects by the magnetic field. On one hand, the external field suppresses the spinon transport; on the other hand, it weakens the scattering between spinon and phonon. Whereas, the broad valley at $\kappa(B)$ reminds us the phonon resonant scattering by free spins \cite{Oxford, NGSO, GBCO, PLCO}.

\section{SUMMARY}

We grew high-quality CeTa$_7$O$_{19}$ single crystals and studied the low-temperature magnetic susceptibility, specific heat and thermal conductivity. The dc magnetic susceptibility and magnetization confirm that it is an effective spin-1/2 system with easy-axis anisotropy and antiferromagnetic spin coupling. The ultralow-temperature ac susceptibility and specific heat data indicate the absence of any phase transition down to 20 mK. The ultralow-temperature thermal conductivity at zero magnetic field exhibits a non-zero residual term $\kappa_0/T =$ 0.0056 W/K$^2$m, while the 14 T thermal conductivity shows an essential zero residual term. All these results suggest a possible ground state of quantum spin liquid.

\begin{acknowledgements}

This work was supported by the National Key Research and Development Program of China (Grant No. 2023YFA1406500), the National Natural Science Foundation of China (Grants No. 12404043, No. 12274388, No. 12474098, No. 12104011, and No. 12204004) and the Nature Science Foundation of Anhui Province (Grant No. 2408085QA024). The work at the University of Tennessee was supported by the NSF with Grant No. NSF-DMR-2003117. H.W. and W.X. are supported by the U.S. DOE-BES under Contract No. DE-SC0023648. A portion of this work was performed at the National High Magnetic Field Laboratory, which is supported by National Science Foundation Cooperative Agreement No. DMR-2128556* and the State of Florida and the U.S. Department of Energy. M. L. and S. Z. acknowledge support from the LDRD program at Los Alamos National Laboratory.

\end{acknowledgements}

\end{document}